\def\comment#1{}
\newcommand{\Tr}{\mbox{Tr\,}}

\newcommand{\tr}{\mbox{tr\,}}
\newcommand{\be}{\begin{eqnarray}}
\newcommand{\ee}{\end{eqnarray}}

\newcommand{\sla}[1]{{\hspace{1pt}/\!\!\!\hspace{-.5pt}#1\,\,\,}\!\!}

\documentstyle[prl,aps,amsfonts,psfig,multicol,epsf]{revtex}

\def\cm#1{}

\begin{document}
\title{{
Mass generation without phase  coherence 
in the Chiral Gross-Neveu Model  \\ at
finite temperature and small $N$ in 2+1 dimensions.
}}
\author{
  Egor Babaev
\thanks{Email: egor@teorfys.uu.se  \ \
 http://www.teorfys.uu.se/people/egor/   \ \
 \ \ Tel: +46-18-4717629  Fax +46-18-533180
}}
\address{
Institute for Theoretical Physics, Uppsala University
Box 803, S-75108 Uppsala, Sweden 
}
\maketitle
\begin{abstract}
The chiral Gross-Neveu model is one of the most 
popular toy models for QCD. In the past, it has been studied in detail 
in the large-N limit. In this paper we study its
small-N behavior at finite temperature in 2+1
dimensions. We show that at small N the phase diagram of this
model is {\it principally} different from  its behavior at $N\rightarrow \infty$.
We show that for a small number $N$
of fermions the model possesses two characteristic 
temperatures $T_{KT}$ and $T^*$. That is, at small N,
along with  a quasiordered phase $0<T<T_{KT}$ 
the system possesses a very large 
region of precursor fluctuations $T_{KT}<T<T^*$ which 
disappear only at a temperature $T^*$,
 substantially higher than the temperature  $T_{KT}$
of Kosterlitz-Thouless transition.
\comment{
corresponds to onset of quasi-long-range order and
$T^* (>T_{KT})$ that corresponds to thermal
breaking of fermion pairs. This phenomenon is {\it not} new but
is known from the recent studies of superconductors.
The region $T_{KT}<T<T^*$ is characterized by
a complex gap function $|\Delta (x)| e^{i \varphi (x)}$ with {
\it  nonzero gap modulus} $|\Delta|$ but random phase $\varphi(x)$
so there is no phase coherence (or no quasi-long-range order in the case
of a purely 2D system) and system behaves like
a gas of non-condensed composite bosons. In superconductivity
this region calls {\it pseudogap phase}.
In this paper we rederive
these results known from
superconductivity in a very simple and trasparent way
in a GN model with $U(1)$-symmetry and discuss
its possible relations to particle
physics. In particular we discuss possibility of
generalization of these results to higher dimensions
and other symmetries
and roles of quantum and classical fluctuations.
In the regime of large $N$ we also show that the temperature
of the phase transition of XY-model tends from below to the temperature
of the pair formation and merges with it in the limit $N \rightarrow
\infty$ thus recovering mean-field scenario for the onset
of quasi-long-range order in this model.}
\end{abstract}
\begin{multicols}{2}
\narrowtext
In this letter we discuss small-N
behavior of the Gross-Neveu (GN)
\cite{GNM} model with U(1)-symmetry
in $2+1$ dimensions at finite temperature.
The Gross-Neveu model 
is a
field theoretic model of zero-mass fermions
with quartic interaction,
which provides
us with considerable insight into the
mechanisms of spontaneous symmetry breakdown
and is considered to be an illuminating toy model for QCD.
Our small-N study is motivated by recent 
results in the theory of superconductivity in the 
regimes when BCS mean-field theory is 
not valid.

Recently a remarkable progress was made
in the theory of superconductivity 
in understanding mechanisms of symmetry breakdown 
in the regimes of strong interaction and 
low carrier density. 
 It was observed that, 
in general, a Fermi system with attraction possess
two distinct characteristic temperatures corresponding 
to pair formation and pair condensation.
That is, in a strong-coupling superconductor 
Cooper pairs are formed at a certain temperature $T^*$
although there is no macroscopic occupation 
of zero momentum level and thus no phase coherence 
and no symmetry breakdown. The temperature 
should be lowered to $T_c (<< T^*)$ in order to
make these pairs condense and establish phase coherence.
The large region $T_c < T < T^*$ where there exist 
Cooper pairs but no continual  symmetry is broken 
is called {\it pseudogap phase}. Thus the symmetry breakdown
in a strong-coupling superconductor resembles 
onset of long range order in ${}^4 He$ where, formally
one also can introduce a characteristic temperature 
of thermal decomposition of a $He$ atom, however
it does not mean that this temperature is connected
in any respect with temperature of the onset 
of phase coherence.
In fact, the BCS scenario 
when superconductive phase transition can
approximately be described as a pair-formation transition
(i.e. when there is only one characteristic
temperature $T_c$)
is a very exceptional case. That is, the BCS scenario
 holds true only at infinitesimally weak
coupling strength or very high carrier density.

Similarly as the BCS theory became a source of inspiration 
in particle physics (in particular it had  
 direct influence on appearance of the  Nambu--Jona-Lasinio
 and Gross-Neveu models), recently the pseudogap concept started
to penetrate from the theory of superconductivity to the 
high-energy physics, sparking many intensive discussions.
The pseudogap concept was first introduced to 
particle physics in \cite{gn1}.

Let us now return to the  Gross-Neveu model.
At finite temperatures
in the limit of large N its behavior closely resembles a BCS
superconductor \cite{park}.
Below, we show that 
a very rich physical structure, similar to  the phase diagram
of a strong-coupling superconductor, emerges
in the  small-N limit in the chiral GN model.

Existence of the  pseudogap phase 
in the chiral GN model at small N provides us with an
example
when generation
of fermion mass happens  without  spontaneous
breakdown of continual symmetry and suggests 
possibility of importance of this concept in particle 
physics.
%
%
%

The chiral GN model \cite{GNM} has the following
 Lagrange density \cite{gam}
\begin{eqnarray} \label{8.67b}
  {\cal L} = \bar\psi_a  i\sla{\partial}
       \psi _a + \frac{g_0}{2N}
\left[
\left( \bar\psi _a       \psi _a\right) ^2
+\left( \bar\psi _a  i \gamma_5     \psi _a\right) ^2
\right]  .
\nonumber
\end{eqnarray}
where the index $a$ runs from $1$ to $N$.
The fields $\psi(x) $ can be  integrated
out yielding collective field action:
\begin{eqnarray} \label{8.74b}
  {\cal A}_{\rm coll} [\sigma ] = {N} \left\{ -
           \frac{\sigma ^2+\pi^2}{2g_0} - i \Tr \log
            \left[ i \sla{\partial}  - \sigma (x)-i \gamma_5\pi\right]
            \right\}.
\nonumber
\end{eqnarray}
This model is invariant under the continuous set of chiral O(2)
transformations which rotate $ \sigma$ and $ \pi$ fields into each other.
In large-N limit,  the model has 
a second order phase transition at which 
 fermions acquire mass. At 
zero temperature in 2+1 dimensions it is accompanied 
by an appearance of a 
 massless composite Goldstone boson 
(for details see e.g. the review \cite{park}).
In the
symmetry-broken phase
the model is characterized by 
a typical ``mexican hat" effective potential.
The propagator of the massive $\sigma$  fluctuations
can be readily extracted 
and it coincides with the $\sigma$-propagator 
of the ordinary GN
model \cite{park,gn1}:
\be
G_{\sigma '  \sigma ' } = \nonumber ~~~~~~~~~~~~~~~~~~~~~~~~~~~~~~~~~~~~~~~~~~~~~~~~~~~~~~~~~~~~~~ \\ 
-\frac{i}{N} \left[
g_0^{-1}
- i  \ \mbox{tr} 
\int \frac{d^3 k}{(2 \pi)^3}
\frac{(\sla{k}+M)( \sla{k}- \sla{q}+M)}
{(k^2 -M^2)[(k-q)^2-M^2]}
\right]^{-1}
\nonumber
\ee
Where $M$ is the mass dynamically acquired by fermions.
At finite temperature according to standard dimensional
reduction arguments the system is effectively two dimensional
and thus the Coleman theorem forbids the spontaneous 
breakdown of the U(1)-symmetry. However, as it was shown 
 by Witten \cite{W},
this does not preclude the  system from generating fermion mass.
That is, as was shown 
in \cite{W},  employing ``modulus-phase" variables
\be
\sigma + i \pi = |\Delta|e^{i \theta}
\label{variables}
\ee
 one  can see that 
the system generates fermion mass $M=|\Delta|$
that coincides with the modulus
of the complex order parameter, but its phase 
remains incoherent  
and  the correlators of the complex order parameter 
have algebraic decay.
Existence 
of the {\it local} gap modulus $\Delta$ does not contradict 
Coleman theorem since $\Delta$ is neutral under $U(1)$ transformations.
Thus at low temperature in 2+1  dimensions 
there appears an ``almost" Goldstone boson
that becomes a real Goldstone boson at exactly zero temperature. 

Let us start our study 
with an inspection of the effective potential of the 
model at finite temperature in the leading order approximation and then 
take into account the next-to-leading order corrections.
Following  \cite{W}, 
the fermion mass at finite temperature
is given by the gap equation which coincides
with the gap equation for ordinary GN model with discrete symmetry
(for detailed calculations see \cite{park,gn1}):
\begin{eqnarray} \label{t2}
1 & = &  g_0     \,
 \tr  (1)\int \frac{d^{2} k}{(2\pi )^{2}}
                    \frac{1}{2E}  \tanh \left(
\frac{E}{2T} \right),
\end{eqnarray}
where $E$ stands for $\sqrt{k^2+\Delta^2}$.
In the leading order mean-field approximation 
we have a situation
similar to BCS superconductor: there is a gap
that disappears at certain temperature
which, in what follows, we denote by
$T^*$.
It can be expressed via the gap function at zero temperature:
%
%
%
\begin{equation}
\label{tstar}
T^* = \frac{\Delta(0)}{2 \log 2},
\end{equation}
Near $T^*$ the gap function has, in the mean-field
approximation, the following behavior:
\begin{equation}
\label{mhigh}
\Delta(T) = T^* 4 \sqrt{\log 2} \sqrt{ 1 - \frac{T}{T^*}}
\end{equation}
On the other hand at low
temperatures the gap function receives exponentially small
temperature correction:
\begin{equation}
\label{mzero}
\Delta(T) =  \Delta(0) -2 T\exp \left(
- \frac{\Delta(0)}{T}
\right)
\end{equation}
Lets us stress once more: a straightforward calculation of 
next-to-leading order 
corrections gives that the gap should be 
exactly zero at any finite temperature in 2+1 dimensions.
However as it was shown in the  
paper  mentioned above \cite{W} such a direct calculation of corrections 
misses essential physics of a two-dimensional problem. 
The fluctuations can be made arbitrarily weak by
decreasing temperature in 2+1 dimensions 
(or e.g. increasing N
in 2D zero temperature  calculations in \cite{W}) 
and then the system possesses a very well-defined ``mexican hat"
effective potential that determines fermion mass.
Due to phase fluctuations in the degenerate valley of the potential, the 
average of the complex gap function is zero, however
there exist a gap locally (i.e. in some sense the system in 
its low energy domain degenerates to a nonlinear sigma model).
In 2+1 dimensions,  as the temperature approaches
zero the thermal fluctuations in the degenerate valley 
of the effective potential gradually vanish and at $T=0$
a local gap becomes a real gap. Most interesting 
effect happens however when temperature is increased.
It was anticipated before that there is only one characteristic
temperature in a such system - namely
 the temperature of the Kosterlitz-Thouless (KT)
 transition which (as was anticipated) coincides with 
temperature of the formation 
of the local gap. We show below that this scenario 
holds true only at very large number of field components. 
In general, as we show,
 the model has two characteristic temperatures like in the 
case of a superconductor with a pseudogap.
In order to show it we have to go beyond mean-field
approximation.

Let us make an expansion around
a saddle point solution and derive the 
propagator of the imaginary part of the
field $\Delta$ that has a pole at $q^2=0$.
The procedure is standard and will not be 
reproduced here, see for details e.g.  \cite{gn1,park}.
%
\begin{eqnarray} \label{stiff}
G_{\Delta{_{\rm im} '}{\Delta{_{\rm im} '}}} =
\frac{1}{N}
\left[
\frac{1}{8 \pi \Delta(T)} \tanh
\left(
\frac{\Delta(T)}{2T}
\right)
\right]^{-1}
\frac{i}{q^2}
\end{eqnarray}
The propagator (\ref{stiff}) characterizes 
stiffness of the phase fluctuations in the degenerate minimum
of the effective potential. 
It gives the following expression for the kinetic term 
of phase fluctuations for the chiral GN model:
\begin{equation}
\label{kin}
E_{kin}= \int d^2 x
\frac{N}{8 \pi} \Delta(T) \tanh
\left(
\frac{\Delta(T)}{2T}
\right)
[\nabla \theta]^2
\end{equation}
Now we have all the tools to find the position of the KT 
transition in the chiral GN model.
It is  well-known, that the KT transition can not 
be found by straightforward perturbative methods.
 In order to find a position 
of the KT transition one should first observe that 
in variables (\ref{variables}) the system is described
by a complex scalar field. The key feature is that 
the field $\theta$ is cyclic: $\theta = \theta + 2\pi n$.
In two 
dimensions such a system possesses  excitations
of the form of vortices and antivortices that are
attracted to each other by a Coulomb potential.
At low temperatures,
the vortices and antivortices
form bound pairs.
The grand-canonical ensemble of the pairs exhibits
quasi-long-range correlations.
At certain  temperature $T_{KT}$,
the vortex pairs break up which is the Kosterlitz-Thouless
phase-disorder transition \cite{bkt}.
This transition was studied in detail 
in the field theory of a pure phase field $\theta(x)$,
with a Hamiltonian
\begin{equation}
{\cal H}=\frac{\beta}{2}[\partial \theta(x)]^2,
\label{@modelld}\end{equation}
where  $\beta$ is the stiffness of the  phase fluctuations.
In our case the coefficient $\beta$ depends on temperature
and parameters of the GN models, namely number of field
components and  $\Delta$ [see eq. (\ref {kin})].

The temperature of the Kosterlitz-Thouless phase
transition  in the system (\ref{@modelld}) is given by \cite{bkt}:
\begin{equation}\label{tkt}
 T_{\rm KT}=\frac{\pi}{ 2} \beta.
\end{equation}
In order to study the phase disorder transition in
the chiral GN model
we should solve a {\it set} of equations:
equation for $T_{KT}(\Delta,N)$ that follows from
the kinetic term and equation (\ref{t2}) for the gap modulus
$\Delta(T_{KT},N)$ that follows from
effective potential. I.e. in our case the phase transition
is a competition of two processes: the thermal excitation of
directional fluctuations in the degenerate valley of the effective
potential and the thermal depletion of the stiffness coefficient.
%
%

Let us first consider the case of small N.
From expressions (\ref{kin}), (\ref{tkt}), (\ref{tstar}), (\ref{mzero})
we see that in the regime of small N \ \ \
$T_{KT} << T^*$. In this regime
the temperature corrections to the
phase stiffness are exponentially suppressed.
Thus at temperatures
$T << T^*$
the asymptotic expression for the  kinetic term (\ref{kin})
reads
\begin{equation}
\label{kinlow}
H_{kin}=  \int d^2 x
\frac{N}{8 \pi} \Delta(0) [
\nabla \theta]^2
\end{equation}
Thus the Kosterlitz-Thouless transition will
take place at the temperature:
\begin{equation}
\label{ktlow}
T_{KT}= \frac{N}{8} \Delta(0)
\end{equation}
which at small $N$ is significantly lower than
the temperature  (\ref{tstar}) 
at which the gap modulus disappears.
For the ratio $T_{KT}/T^*$ at small N
we obtain:
\begin{equation}
\label{ktlow1}
\frac{
T_{KT}}{T^*}= \frac{N \log (2)}{4}
\end{equation}
So with decreasing N separation of $T_{KT}$ and $T^*$
increases.
%
Lets us now turn to the regime when $N$ is no longer small.
Then from the equations
(\ref{kin}), (\ref{tkt}), (\ref{tstar}), (\ref{mhigh})
we see that in this regime
$T_{KT}$ tends from below to $T^*$. The
Hamiltonian (\ref{kin}) in this limit reads near $T^*$:
\begin{equation}
\label{kinhigh}
H_{kin}= \int d x
\frac{N}{16 \pi} \frac{ \Delta(T)^2}{T}
[\nabla \theta]^2
\end{equation}
From the eqs. (\ref{kinhigh}), (\ref{tkt}), (\ref{tstar}), (\ref{mhigh})
we find the  following expression
for behavior of $T_{KT}$
at large N:
\begin{eqnarray}
\label{merger}
T_{KT} 
\simeq
 T^* \left(1 - \frac{1}{N \log (2)} \right).
\end{eqnarray}
This equation explicitly shows merger of the temperatures
 $T_{KT}$ and $T^*$ in the limit of large N. This 
can be interpreted as the restoration of the ``BCS-like"
scenario for the quasi-condensation in the limit
$N \rightarrow \infty$.
%
The ratio $T_{KT}/T^*$ is displayed on Fig.~1.
\begin{figure}[htpb]
\epsfxsize=0.7\columnwidth
\centerline{
\epsffile{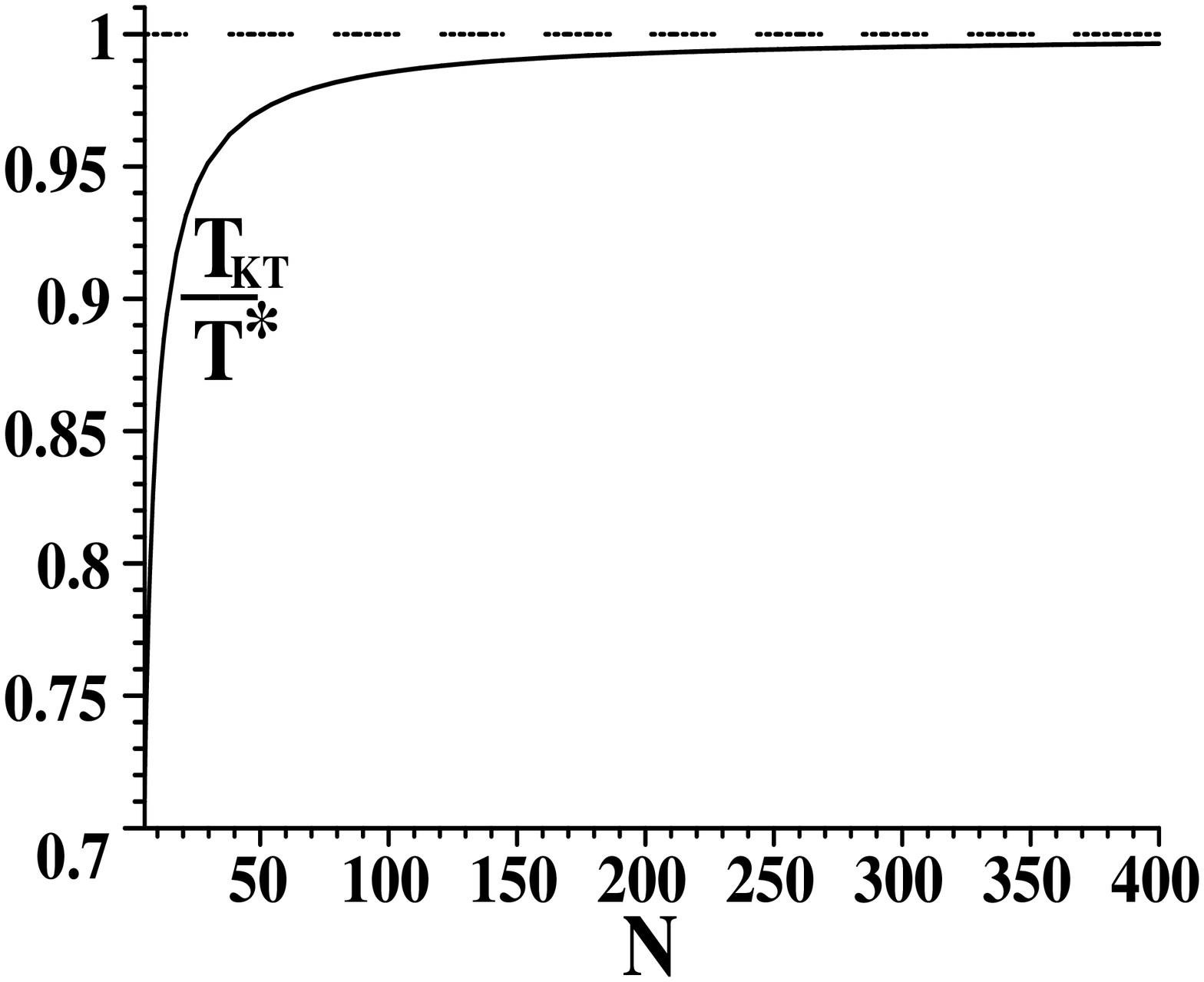}
}
\vskip 0.3cm
\caption{Recovery of a ``BCS-like" scenario 
for quasicondensation at large N in the chiral GN model.
The solid curve is the ratio
of the temperatures of the KT transition ($T_{KT}$) and the characteristic
temperature of the formation of the effective potential ($T^*$).
As it is shown on the picture this ratio tends from below to unity
(the horizontal dashed line)  as N is increasing 
and the region of precursor fluctuations
shrinks.
}
\end{figure}

\comment{
\begin{figure}[tbh]
\input Plot.tps
\narrowtext\vspace{2mm}
\caption{ Ratio
of the temperature of the Kosterlitz-Thouless transition to the
temperature of the gap modulus formation
$T_{KT}/T^*$ in GN model with U(1) symmetry
in the regime of {\it high} $N$.
These temperatures merge in the limit $N \rightarrow \infty$
thus recovering mean-field scenario for the transition
into the quasi-ordered state. In the opposite limit
of {\it small} $N$ separation of these temperatures
is given by formula: $T_{KT}/T^*  = N \log(2)/8 \approx 0.087 N $}
\label{f1}\end{figure}}

%
Thus the phase diagram of the model at {\it small } N consists
of the following phases at non-zero temperature:
(i) 
 $0<T < T_{KT} $ -
the low temperature quasi-ordered phase
with bound vortex-antivortex pairs.
(ii) $T_{KT} < T < T^* $ - 
the phase analogous to {\it pseudogap phase }
of superconductors: i.e. the chirally symmetric phase 
with unbounded vortex-antivortex pairs
that exhibit violent precursor fluctuations and
a nonzero local modulus of the complex gap function.
(iii)  $T>T^*$ - high temperature ``normal"
chirally  symmetric phase.

The mechanism of the phase separation is very transparent 
with the key feature being the  fact that the stiffness
is proportional to $N$ [see eq. (\ref{kin})]. At large $N$ the 
directional fluctuations are energetically extremely expensive and thus, 
the phase transition is controlled basically by the modulus 
of the order parameter. On the other hand at small $N$,
the stiffness is low, and
the thermal excitation of  the fluctuations in the degenerate 
valley of the effective potential starts 
governing the phase transition in the system.

Now let us briefly discuss the physical meaning of $T^*$
and what one can expect
to happen  when the system reaches it at small N. 
At first, as we can conclude from  simple 
physical reasoning in analogy with superconductivity, the 
appearance of the second characteristic temperature 
is very natural.
Besides the fact that the phase analogous to the 
intermediate phase $T_{KT}<T<T^*$ it is the dominating region on a phase
diagram of strong-coupling and low carrier density superconductors,
the similar effects are known 
in a large variety of condensed matter systems such as 
exitonic condensates, Josephson juncion arrays and many other
systems.
One of the most  illuminating examples of
the appearance of the pseudogap phase 
is the chiral GN model  in $2+\epsilon$
dimensions
at zero temperature where this phenomenon 
is governed by quantum dynamical fluctuations at small $N$ \cite{gn1}.
In D=$2+\epsilon$ the presence of two small parameters in the 
problem has allowed to prove existence and
different physical origin of two characteristic values of
renormalized coupling constant and the formation of an intermediate 
pseudogap phase \cite{gn1}. 
We can also observe that the mean-field 
approximation gives a second order phase transition at $T^*$
which is certainly an artifact
 since much above $T_{KT}$ there
are violent thermal phase fluctuations.
These fluctuations should wash out 
the phase transition at $T^*$ which should degenerate to a smeared 
crossover as it happens in superconductors. Apparently, this
crossover can not be studied adequately 
in the framework of $1/N$-expansion.
The most insight into the properties of the system in the 
region $T_{KT}<T<T^*$ can be obtained by numerical simulations.
Although the KT transition is very hard to observe in simulations,
the hint for the phase separation would be a gradual degradation 
of the transition at $T^*$ with decreasing $N$ .

Important circumstance is that the pseudogap phase as
a precursor of the chiral phase transition is not  
a phenomenon common only for low dimensional systems but also
should occur at small N in higher dimensions and 
also in systems with other symmetries. 
One should however observe that as  it
was shown recently by us,  unfortunately 
no {\it direct} generalization of the discussed
here nonlinear sigma model approach  is possible 
to NJL model in $3+1$ dimensions \cite{NJ}, at least in a closed form. 

Unfortunately  
the letter format of this paper does not allow us 
to discuss in detail the similarities and differences
of these phenomena in the chiral GN model
and in the models of superconductivity with 
precursor formation of Cooper pairs. 
A detailed discussion of these aspects is
currently in preparation and will be presented elsewhere.
\comment{
The described situation is similar
to strong-coupling or low carrier density behavior
of a superconductor in two and three dimensions
in which case the region where pairs are
formed but not condensed is called {\it pseudogap phase}
(see \cite{sc} and references therein).
This phenomenon is observed experimentally \cite{u}.
A similar situation was found as well in our previous paper
on condensate depletion by {\it quantum}
fluctuations in the Chiral GN model in $2 + \epsilon$
dimensions at small N.}
%
%

The author is grateful to Dr. V. Cheianov, 
Prof. H. Kleinert and Prof. A.J. Niemi for discussions.

\end{multicols}

\begin{thebibliography}{999}
%
\bibitem{GNM}
{D.\ Gross} and {A.\ Neveu}, Phys.\ Rev.\ D {\bf 10} 3235 (1974).

\bibitem{gam}
In $2+1$ dimensions we choose $\gamma$-matrices as in the 
review \cite{park}:
 $\gamma^{\mu} = \sigma^{\mu} \otimes
\left(\begin{array}{cc}
I & 0 \\
0 & -I
\end{array} \right)$ and $i \gamma_5 =
\left(\begin{array}{cc}
0 & I \\
-I& 0
\end{array} \right)$.

\bibitem{park}
B. Rosenstein, B.J. Warr, S.H. Park
Phys. Rep. {\bf 205} 59 (1991)
\bibitem{W} E.~Witten, Nucl.~Phys.~B {\bf 145}, 110 (1978)
\bibitem{gn1}
H. Kleinert and E. Babaev  preprint hep-th/9809112;
Phys.Let. B. 438, 311 (1998). [There is a misleading 
statement in the above paper about a ``proliferation" 
transition in 3D].
\bibitem{bkt}
V. L. Berezinskii.
Zh.~Eksp.~Teor.~Fiz., 1970, { \bf 59} 907;
J. Kosterlitz, D. Thouless.
J.~Phys., 1973, {\bf C6} 1181.
\comment{
\bibitem{gap}
The same assumption in the
case of the chiral Gross-Neveu model was made implicitly
when we derived the gap function in the mean-field approximation
i.e. we
postulated that phase is constant and determined the
gap size with a standard variation procedure.}
\bibitem{NJ}
E. Babaev preprint hep-ph/0006087 [accepted to Phys. Rev. D].
In this paper we also criticize another recent attempt of introduction of a
nonlinear sigma for description of chiral fluctuations
in the NJL in 3+1 dimensions made in 
H. Kleinert and B. Van den Bossche Phys.Lett. B474 (2000) 336. 
\end{thebibliography}
\end{document}